\documentclass[usegraphicx]{mn2e}
\usepackage{longtable}
\usepackage{amsmath,amssymb}
\usepackage{times}
\usepackage{subfigure}
\usepackage{epsfig}
\usepackage{graphics}

\renewcommand{\descriptionlabel}[1]%
  {\hspace{\labelsep}\textbf{#1}}

\title[Period changes of RR Lyrae stars in M5]
     {A New Overview of Secular Period Changes of RR Lyrae stars in M5}

\author[A. Arellano Ferro et al.]
{A. Arellano Ferro$^{1}$, J. A. Ahumada$^{2}$, N. Kains$^{3}$, A. Luna$^{1}$\\
$^{1}$Instituto de Astronom\1a, Universidad Nacional Aut\'onoma de M\'exico.
Ciudad Universitaria CP 04510, Mexico: (armando@astro.unam.mx)\\
$^{2}$Observatorio Astron\'omico, Universidad Nacional de C\'ordoba, 
Laprida 854, 5000 C\'ordoba, Argentina: (javier@oac.unc.edu.ar)\\
$^{3}$Space Telescope Science Institute, 3700 San Martin Drive,
Baltimore, MD 21218, United States of America: (nkains@stsci.edu) \\
}

\begin{document} 

\date{Accepted -----. Received ----; in original form ----}

\pagerange{\pageref{1}--\pageref{15}} \pubyear{2016}

\maketitle 

\label{firstpage}

\begin{abstract}
Secular period variations, $\beta=\dot{P}$, in 76 RR Lyrae stars in the globular
cluster M5 are analysed
using our most recent CCD $V$ photometry and the historical photometric database
available in
the literature since 1889. This provides a time baseline of up to 118 years for these
variables.
The analysis was performed using two independent approaches: first, the classical
$O-C$
behaviour of the time of maximum light, and second, via a grid $(P,\beta)$, where the
solution
producing the minimum scatter in the phased light curve is chosen. The results of the
two
methods agree satisfactorily. This allowed a new interpretation of the nature of the
period
changes in many RR Lyrae stars in M5. It is found that in $96\%$ of the stars studied
no irregular
or stochastic variations need to be claimed, but that $66\%$ of the population shows
steady period
increases or decreases, and that $34\%$ of the periods seem to have been stable over
the last century.

The lack of systematic  positive or negative period variations in RR Lyrae stars in
other globular clusters is addressed, and the paradigm of period changes being a
consequence of stellar evolution is discussed in terms of chemical variations near the
stellar core and of multiple stellar populations. In M5 we found a small positive
average value of $\beta$ and a small surplus of RRab stars with a period increase.
Nevertheless, in M5 we have been able to isolate
a group of likely evolved stars that systematically show positive,
and in some cases large, period change rates.

\end{abstract}

\begin{keywords}
globular clusters: individual (NGC~5904) -- stars:variables: RR Lyrae
\end{keywords}

\section{Introduction}

Secular period changes in the RR Lyrae stars (RRLs) of NGC 5904 (M5, or C1516+022 in
the IAU nomenclature) ($\alpha = 15^{\mbox{\scriptsize h}}
18^{\mbox{\scriptsize m}} 33.2^{\mbox{\scriptsize s}}$, $\delta = +02\degr 04\arcmin
51.7\arcsec$, J2000; $l = 3.85\degr$, $b = +46.80\degr$), have been measured
since the pioneering work of Oosterhoff (1941), and in the following seven decades 
several authors have tackled the problem of estimating the period change rates
in pursuit of measurable indications of the stellar evolution on the horizontal branch
(HB) accross the instability strip (IS) (Coutts \& Sawyer-Hogg 1969, Kukarkin \&
Kukarkina 1971, Storm et al. 1991, Reid 1996, Szeidl et al. 2011). Photometric
studies of M5 are numerous, as
are time-series data  of its variable stars since 1889 (Bailey 1917), leading to a
wealth of resources available for the study of secular period variations.

The most recent and complete study to date of the period variations in the RRLs in
M5 was published by Szeidl et al. (2011) (herafter SZ11), who employed a
new approach to the $O-C$ method that uses complete light curves rather than a
single specific phase. They reported period changes for 86 stars, using a time
baseline of about 100 years for the majority of them. While many stars were found to
have a uniform
period increase or decrease, several very intriguing cases of irregular
period variations emerged in that study which are difficult to confront with
theoretical
expectations. In the present study we have added our most recent CCD time-series
of M5 (Arellano Ferro et al. 2016a, hereafter AF16) that extends the time baseline, in
many cases, to about 118 
years, which have allowed us to re-interpret some of the $O-C$ diagrams,
particularly the ones of those peculiar stars.

The present paper is organised as follows; in $\S$ \ref{Previous} we summarise the
previous studies of period changes in M5
and the available data collections in the literature. In $\S$ \ref{Methods}
the methods employed in this work to calculate period change rates are described,
the results are reported and a brief discussion of particular cases is included.
$\S$ \ref{EVOL_HB} contains a discussion on the lack of systematic period
change rates as a consequence of several effects, other than simple evolution,
that influence the
stellar structure and chemistry and hence the period change rates. A summary of our
results and conclusions is given in $\S$ \ref{summary}.

\section{Observations}

The most recent $V$ CCD time-series used in this paper to substantially extend the
time baseline in many cases to about 118 years, was performed on 11 nights between
February 29, 2012 and April
09,
2014 with the 2.0~m telescope at the Indian Astronomical Observatory (IAO), Hanle,
India, and have been reported by AF16. A total of
392 images obtained in the Johnson-Kron-Cousins $V$ filter are used here for the
purpose of the present analysis. For a full description of observations and the
reduction procedure, the reader is referred to AF16.

\begin{table}
\caption{Sources of photometric data and times of maximum light in the
RR Lyrae stars of M5.}
\centering
\begin{tabular}{lccccc}
\hline
Authors  &  years & band & notes  \\
\hline
Bailey (1917)&1889-1912&$pg$ &1 \\
Shapley (1927) &1917& $pg$& 2 \\
Oosterhoff (1941)     & 1932-35& $pg$& 3 \\
Coutts \& Sawyer-Hogg (1969) &1936-66& $pg$ &  \\
Szeidl et al (2011) &1952-1963 & $pg$& \\
Szeidl et al (2011) &1971-1993 & $pg$& \\
Kukarkin \& Kukarkina (1971) & & &4\\
Cohen \& Gordon (1987) &1986& $B,i$ &  \\
Storm et al. (1991) &1987& CCD $V$ &  \\
Cohen \& Matthews (1992) &1989& CCD $V$ &  \\
Brocato et al. (1996) & 1989 & CCD $V$ &\\
Reid (1996) & 1991-92&CCD $V$& \\
Kaluzny et al. (1999) & 1997& CCD $V$ &  \\
Present work & 2012-14& CCD $V$ &  \\
\hline
\end{tabular}
\label{archDATA}
\raggedright
\center{\quad 1. Times of maximum light were calculated by Kukarkin \& Kukarkina
(1971), 2: Light curves and times of maximum in
Coutts \& Sawyer-Hogg (1969) and Coutts (1971). 3: Times of maximum in
Coutts \& Sawyer-Hogg (1969). 4: No new data are provided but it is a source of times
of
maximum light.}
\end{table}

\begin{table}
\footnotesize
\caption{Observed times of maximum light $O$ for RR Lyrae stars in M5 and the
corresponding $O-C$ residuals calculated with the given ephemerides for each variable.
The sources of either the times of maximum light or the data employed to calculate the
times of maximum are coded in column 4 as follows: KK (listed by Kukarkin \& Kukarkina
1971); hWil (calculated in this work from Mount Wilson 1917 data published by Coutts
\&
Sawyer-Hogg 1969); hDDO (calculated in this work from DDO data published by
Coutts \& Sawyer-Hogg 1969);
CSH (listed by Coutts \& Sawyer-Hogg 1969 from Oosterhoff's 1941 data); COU (listed
by Coutts 1971 from Mount Wilson 1917 data); hOOS (calculated in this work from
Oosterhoff's 1941 data); 
STO (calculated in this work from data by Storm et al. (1991);
BRO (calculated in this work from data by Brocato et al. 1996); KAL (calculated
from data by Kaluzny et al. 1999), B24, A49, L24
and  P40 (calculated from the data in Table 1 of SZ11); COH (calculated
in this work from data by Cohen \&
Sarajedini 2012); AF16 (calculated from data by AF16).
This is an extract from the full table, which is only available with the electronic
version of the
article (see Supporting Information).}
\centering
\begin{tabular}{lclc}
\hline
Variable&$P_0 (days)$ & $E_0$ (HJD) & \\
\hline
V1& 0.567513&2456504.1766&\\
\hline
$O$ (HJD)&$(O-C)$ & No. of Cycles & source\\
240 0000.+  &  days &            &  \\
\hline
13715.5848&+0.6066  &--82004.& KK  \\
16250.4334&+0.5714  &--77146.& KK  \\
19535.6214&+0.5443  &--70850.& KK  \\
21350.3764&+0.4988  &--67372.& KK  \\
21375.9511&+0.5069  &--67323 & hWil \\
21424.9782&+0.4883  &--67229.& COU  \\
27563.8074&+0.4078  &--55464 & CSH \\ 
27610.2469&+0.4092  &--55375.& KK  \\
33858.6369&+0.3190  &--43400 & hDDO \\
34154.497 &+0.3228  &--42833.& B24  \\
36762.3805&+0.2774  &--37835.& KK  \\
38071.5434&+0.2613  &--35326.& KK  \\
39065.5396&+0.2457  &--33421.& KK  \\
39942.1631&+0.2458  &--31741.& KK  \\
41007.6310&+0.2144  &--29699.& A48  \\
42126.8660&+0.2013  &--27554.& L24  \\
42954.4190&+0.1890  &--25968.& P40  \\
50578.7717&+0.0878  &--11356.& KAL  \\ 
56063.2547&--0.0059    &--845.&AF16  \\
56504.1766&0.0      &  0.    &AF16  \\
56757.2390&--0.0077  &+485. &AF16  \\
\hline
\hline
\end{tabular}
\label{tab:maxima}
\end{table}

\section{Previous studies of RR Lyrae secular period variations in M5}
\label{Previous}

 M5 is one of the most observed globular clusters, and observations since 1889
available in the literature are summarised in Table 1. Including our observations, the
baseline spans 118 years, allowing for a detailed analysis of secular period changes
of the variables in this cluster. We have digitized all available data listed in Table
1 and, since these older data may be useful for future studies, but time-consuming to
put together, we have uploaded them onto the Strasbourg Astronomical Data Center
(CDS).

The first-ever study of period variations of RRLs in M5 was performed by
Oosterhoff (1941) who, with a time baseline of about 40 years and despite
having
only three epochs available, 1895-97, 1917 and 1934, was able to detect the period
variations in 43 RRab stars (RRc stars were not studied) and
to conclude that there was no statistical evidence for preferential
period increases in M5.
Coutts \& Sawyer-Hogg (1969) revisited the cluster adding some 40 years to the
time baseline and also concluded that no systematic period variations in the
cluster could be argued and that the small period change rates found by them
might not necessarily be the result of stellar evolution. To establish this, and
considering the time spent by a star on the HB as RR Lyrae, they estimated that the 
time baseline needed to be extended yet another 30 years.
Kukarkin \& Kukarkina (1971)
extended the period changes analysis for 51 stars adding observations of the
Crimean Station of the Sternberg Astronomical Institute including data from 1952
and 1958-1968. They did not publish their observations but listed series of times
of maximum light for all the stars in their sample and offered $O-C$ diagrams. They
concluded that 27 stars have period increases, 14 period decreases and for 9 the
period remains constant. Further period analyses were carried out by Storm et al.
(1991)
for 10 stars and by Reid (1996) for 30 variables, extending their data time baselines
to the 1980s. The most recent and complete analysis of period changes in M5 was
published by SZ11, who used a time baseline of about 100 years and supplemented the
available data with archival data for the
years 1952-1993 not included before.
One of the most striking results of this study based on a large sample of 
86 RRLs is the large number of
variables exhibiting irregular period variations: 50\% of the RRc and 34\% of the
RRab according to these authors. To explain such period variations would be a
challenge for
the theory of stellar pulsation and evolution across the IS. SZ11 also
concluded that there are no preferential period
increases or decreases, hence no preferential evolution to the red or to the blue.
In the present paper we analyse the period secular variations in a sample of
76 RRLs using two independent methods and revisit the discussion of whether the
period variations are the consequence of stellar evolution alone in
the HB.

\section{Present approaches to the period change rate calculation}
\label{Methods}

We have analysed the period changes in the population of RRLs in M5 by two
methods; a) via the classical $O-C$ diagrams constructed from a fixed given
ephemeris
to predict the times of maximum light and compare them with the observed ones. These
diagrams suggest simple period adjustments if the $O-C$ residuals describe a straight
line, or allow the determination of the period change rate $\beta=\dot P$,
if the residuals describe a parabola or a higher order polynomial. b) Searching on
a grid $(P,\beta)$, for the pair that
produces the light curve with the minimum dispersion. These approaches are described
in detail in $\S$ \ref{OmC} and $\S$ \ref{PBeta} respectively.

\begin{table*}
\caption{New periods and period change rates for RR Lyrae stars in M5. Numbers
in parentheses in column 7 are the uncertainties in the last decimal
place and correspond to the uncertainty in coefficient $A_2$ in eq. \ref{BETA_0}.
Values of $\beta_0$ in boldface are the recommended by our analysis. }
\label{variables}
\centering
\begin{tabular}{clcccccccc}
\hline
Variable& Variable & $P_0$& $E_0$ & $E_0 + A_0$& $P_0 + A_1$  &
$\beta_0$&$\beta_0$&$P_0$&notes\\
Star ID & Type     & (AF16)   & (+2~450~000) &(+2~450~000) &   & $(O-C)$&$(P,\beta)$
&$(P,\beta)$
&\\
        &          & (days)   &   (HJD)           &(HJD) & (days)& (d Myr$^{-1}$)&(d 
Myr$^{-1}$)&(days) & \\
\hline
V1 & RRab Bl & 0.521794 &  6757.2390 &6757.2354& 0.52178651
&{\bf --0.001(3)}&+0.02&0.521787&$b$\\
V3 & RRab    & 0.600189 &  6029.2645 &6029.2694& 0.60018771 &{\bf +0.040(9)}
&0.0&0.600186&\\
V4 & RRab Bl & 0.449647 &  6046.2142 &6046.2148& 0.44964950&--0.084(23) &
{\bf --2.31}&0.449556&$a$\\
V5 & RRab    & 0.545853 &  6061.3453 &6061.3512& 0.54584759&{\bf --0.306(12)} &
--&--&\\
V6 & RRab    & 0.548828 &  5989.4522 &5989.4478& 0.54882336&{\bf --0.113(13)} &
--0.07&0.548826&\\
V7 & RRab    & 0.494413 &  5989.4323 &5989.4383& 0.49441262&{\bf +0.227(3)} &
+0.25&0.494415&\\
V8 & RRab Bl & 0.546251 &  5989.3293 &5989.3665& 0.54625752&+0.474(50) &
+0.12:&0.546240&$a$\\
V9 & RRab    & 0.698899 &  6063.4105 &6063.4203& 0.69889415&{\bf --0.030(13)}
&--0.04&0.698893&\\
V11& RRab    & 0.595897 &  6046.2751 &6046.2751&0.59589193 &+0.009(8) &
{\bf 0.0}&0.595893&$b$\\
V12 & RRab   & 0.467699 &  6046.2434 &6046.2467& 0.46769709&{\bf --0.230(3)} &
--0.28&0.467695&\\
V13 & RRab Bl& 0.513133 &  6046.2579 &6046.2299& 0.51312442&+0.018(55)&
+0.08&0.513127&$b$\\
V14 & RRab Bl& 0.487156 &  6063.2014 &6063.2048&0.48715409 &-- &
--&--&$a$\\
V15 & RRc    & 0.336765 &  6061.4425 &6061.4579&0.33677029 &{\bf --0.043(1)}&
+0.03&0.336768&$a$\\ 
V16 & RRab   & 0.647632 &  6029.1907 &6756.4814& 0.64763260&{\bf +0.064(11)} &
+0.11&0.647635&\\
V17 & RRab   & 0.601390 &  6312.5181 &6312.5207& 0.60138522&--0.204(8): & --&--&$a$\\
V18 & RRab Bl& 0.463961 &  6757.2450 &6757.2504& 0.46395054 &--0.007(29)  &--&--&$b$\\
V19 & RRab Bl& 0.469999 &  6046.2639 &6046.2800& 0.4699988&{\bf +0.111(4)}&--&--&$a$\\
V20 & RRab   & 0.609473 &  5987.4631 &5987.4773&
0.60947601&--0.021(6)&--0.06&0.609473 &$b$\\
V24 & RRab Bl& 0.478439 &  6029.3187 &6029.3310& 0.47843694&{\bf --0.327(11)}
&--&--&$a$\\
V25 & RRab   & 0.507525 &5987.3835&5087.3861&0.50756321&{\bf +0.933(94)}&
+4.01&0.507685&$a$\\
V26 & RRab Bl& 0.622561 &  5989.4356 &5989.4459& 0.62256400  &--0.001(20) &
{\bf 0.0}&0.622564 &$b$\\
V27 & RRab   & 0.470322 &  6063.3866 &6063.3959&0.47033603 &{\bf --0.301(6)}&$-$&--&\\
V28 & RRab Bl& 0.543877 &  6312.5122 &--&-- &--&--0.16: &0.543922&$a$\\
V30 & RRab   & 0.592178 &  5987.4673 &5987.4708& 0.59217587  &--0.007(21) &
{\bf 0.0}&0.592176&$b$\\
V31 & RRc bump   & 0.300580 &  6046.2639 &6046.2466& 0.30058184  &{\bf--0.012(1)}
&--0.05& 0.300580&$a$\\
V32 & RRab   & 0.457785 &  5989.4255 &5989.4275& 0.45778643  &--0.004(4)&
--&--&$b$\\
V33 & RRab   & 0.501481 &  5989.3506 &5989.3502& 0.50148223 &{\bf +0.068(7)}&
+0.09&0.501481&\\
V34 & RRab   & 0.56810   &  6061.2497 &6061.2499& 0.56814325
&0.003(9)&{\bf 0.0}&0.568145&$b$\\
V35 & RRc bump   & 0.308217 &  6063.2209 &--&--&--&--&--&$a$\\
V36 & RRab   & 0.627725 & 6061.1878 &6061.1759&0.62772336&{\bf +0.038(12)}&--&--&$a$\\
V37 & RRab   & 0.488801&6029.3187&6029.3184&0.48880064&{\bf +0.089(3)} &+0.12&0.488802
&\\
V38 & RRab   & 0.470422 &  6504.2422 &6504.2582& 0.47042828 &--0.018(49)&--&--&$a,b$\\
V39 & RRab   & 0.589037 &  6063.3327 &6063.3305& 0.58904270&{\bf +0.081(6)}&
+0.11&0.589044 &\\
V40 & RRc Bl & 0.317327 &  6063.1929 &6063.2066&0.31732968&--0.005(5)&
--0.01&0.317329&$b$\\
V41 & RRab   & 0.488572 &  6061.3027 &6061.3054&0.48856722&{\bf --0.078(12)}&
--0.04&0.488569&$a$\\
V43 & RRab   & 0.660226 &  6029.2970 &6757.4952&0.66022986&{\bf +0.023(11)}&
0.0&0.660229&\\
V44 & RRc    & 0.329599 &  5987.4916 &5987.4848& 0.32960270 &+0.006(12)&
{\bf 0.0}&0.329603&$b$\\
V45 & RRab Bl& 0.616636 &  6061.3946 &6061.3870&0.61663865&{\bf +0.027(5)}&
+0.01&0.616638&\\
V47 & RRab   & 0.539730 &  6757.2200 &6312.4917& 0.53972813 &0.022(12)&{\bf 0.0}&
0.539728&$b$\\
V52 & RRab   & 0.501541 &  6029.2645 &6029.2705& 0.50154260 &{\bf 
+0.044(23):}&--&--&$a$\\
V53 & RRc    & 0.373519 &  5987.5209 &5987.5172& 0.37351709  &{\bf 0.0} &--&--&$a,b$\\
V54 & RRab   & 0.454115 &  5989.3176 &5989.3192& 0.45411516 &{\bf +0.045(7)}
&--&--&$a$\\
V55 & RRc Bl & 0.328903 &  6504.1925&6504.1933&0.32890264&{\bf+0.058(8)}&
+0.07&0.328903&\\
V56 & RRab Bl& 0.534690 &  6061.3186 &6029.2424& 0.53469980&{\bf +0.166(16)}&
+0.20&0.534701&$a$\\
V57 & RRc  & 0.284697 &  6046.2434 &6046.2467&0.28469423&{\bf --0.094(14)}&
--0.01&0.284689&\\
V59 & RRab   & 0.542025 &  6061.2807 &6061.2820&0.54202707&{\bf +0.016(3)}&
--0.03&0.542024&\\
V60 & RRc bump   & 0.285236 &  6504.1518&6504.1550&0.28523571&{\bf --0.018(6)}&
--0.04&0.285235&\\
V61 & RRab   & 0.568642 &  6061.4250 &6061.4286&0.56864272&{\bf +0.266(5)}&
+0.25&0.568642&\\
V62 & RRc bump   & 0.281417 &  5989.3176 &5989.3057&0.28142735&{\bf +0.200(9)}&
--&--&$a$\\
V63 & RRab Bl& 0.497686 &  6756.3534 &6046.1731&0.49768368&{\bf +0.065(9)}&
+0.06&0.497683&\\
V64 & RRab   & 0.544489 &  6062.1833 &6062.1804&0.54449142&{\bf --0.161(27)}&
--0.13&0.544493&\\
V65 & RRab Bl& 0.480664 &  5989.4389 &5989.4474& 0.48067521 &+0.214(7): &--&
0.480672&$a,b$\\
V74 & RRab   & 0.453984 &  6061.3915 &6061.3934& 0.45398467&{\bf --0.119(3)}&
--0.12&0.453985&\\
V77 & RRab   & 0.845158 &  6061.3518 &6061.3484& 0.84514463&{\bf +0.340(21)}&
+0.32&0.845145&\\
V78 & RRc    & 0.264820 &  5989.5204 &5989.5341& 0.26481743  &--0.003(3) &
0.0&0.264817&$b$\\
V79 & RRc    & 0.333139 &  6046.2326 &6046.2411& 0.33313845  &--0.012(4) & 
--&--&$b$\\
V80 & RRc    & 0.336542 &  6046.2751 &6046.2843& 0.33654488&{\bf +0.123(7)}
&--&--&$a$\\

\hline
\hline
\end{tabular}
\raggedright
\center{\quad $^{a}$: There is a comment for this star in $\S$ \ref{sec:IND_STARS}. A
colon after the value of $\beta$ indicates particularly uncertain values.\\
$^{b}$: All these stars display a linear $O-C$ diagram and in fact $\beta$ is
not significantly different from zero. Therfore, we recommend a value of $\beta$ =
0.0}
\end{table*}

\begin{table*}
\addtocounter{table}{-1}
\caption{Continued}
\label{variablesB}
\centering
\begin{tabular}{clcccccccc}
\hline
Variable& Variable & $P_0$& $E_0$ & $E_0 + A_0$& $P_0 + A_1$  &
$\beta_0$&$\beta_0$&$P_0$&notes\\
Star ID & Type     & (AF16)   & (+2~450~000) &(+2~450~000) &   & $(O-C)$&$(P,\beta)$
&$(P,\beta)$
&\\
        &          & (days)   &  (HJD)       &(HJD) & (days)& (d Myr$^{-1}$)&(d 
Myr$^{-1}$)&(days) & \\
\hline
V81 & RRab   & 0.557271 &  6504.2422 &6504.2393& 0.55726945 &{\bf --0.562(9)}&
--&--&\\
V82 & RRab   & 0.558435 &  6063.2014 &6063.2038&0.55843655&--0.051(91)&
--0.23&0.558433&$b$\\
V83 & RRab   & 0.553307 &  6061.4320 &6061.4372&0.55330724&--0.008(6)&--&--&$b$\\
V85 & RRab Bl& 0.527535 &6061.3804&6061.3804&0.52752587&
--0.234(12)&+0.71&0.527537&$a$\\
V86 & RRab   & 0.567513 &6504.1925&6504.1916&0.56751027& --0.065(9)&--&--&$a$\\
V87 & RRab   & 0.738421 & 6061.2186&6061.2050&0.73842323&{\bf +0.369(14)}& +0.35&
0.738423&\\
V88 & RRc    & 0.328090 & 5989.5102&5989.5271&0.32808564&--0.024(21)&--0.45&
0.328080&$b$\\
V89 & RRab & 0.558443 & 6063.1970 &6063.1985&0.55844471&{\bf +0.060(58)}&--0.04&
0.558445&\\ 
V90 & RRab & 0.557168 &  6061.3518&6061.3482&0.55716550&{\bf +0.114(8)}
&+0.10&0.557165&\\
V91 & RRab & 0.584945 &  6063.4183&6063.4085& 0.58494285&+0.011(27)&{\bf 0.0}
&0.584943&$b$\\
V92 & RRab Bl& 0.463388 & 6061.1878&6061.1781&0.46338677 &+0.025(19)&--0.87:
&0.463351&$a,b$\\
V93 & RRab & 0.552300 & 5987.4505&--&--&--  &--&--&$a$\\
V94 & RRab &  0.531327 &  6061.2497 &6061.2475&0.53132796&0.0&--&--&$a,b$\\
V95 & RRc bump & 0.290832 & 6061.3613 & -- &0.29083223&0.0
&+2.15&0.290833&$a,b$\\
V96 & RRab & 0.512255 &  6312.48 &--& 0.512255&0.0&+0.04&0.512255&$a,b$\\
V97 & RRab Bl & 0.544656 &  5987.3747 &5987.3660&0.54464622&--0.134(35)
&+0.85:&0.544654&$a$\\
V98 & RRc & 0.306360 & 6063.4216 &6063.4111&0.30636289&+0.022(14)&--&--&$a$\\
V99 & RRc Bl & 0.321336 & 6061.2186&6061.2059&0.32134529&{\bf +0.159(28)}
&--0.03&0.321344&$a$\\
V100 & RRc & 0.294365 & 6504.1849 &9395.2457&0.29436548&0.0&--&--&$a,b$\\
\hline
\hline
\end{tabular}
\center{\quad $^{a}$: There is a comment for this star in $\S$ \ref{sec:IND_STARS}. A
colon after the value of $\beta$ indicates particularly uncertain values.\\
$^{b}$: All these stars display a linear $O-C$ diagram and in fact $\beta$ is
not significantly different from zero. Therfore, we recommend a value of $\beta$ =
0.0}
\end{table*}

\subsection{The $O-C$ method}
\label{OmC}

The observed minus the calculated ($O-C$) residuals of a given feature in the light
curve, as an indication of miscalculations or authentic variations of the
pulsation or orbital period, 
using a single given phase of the light curve as a reference, is a standard approach
that has been in use for many decades for example in Cepheids, RR Lyrae and contact
binary stars (e.g. Arellano Ferro et al. 1997; Coutts \& Sawyer-Hogg 1969;
Reid 1996; Lee et al. 2011). It is convenient to select a feature that
facilitates the accurate determination of the phase. For RRab stars a good selection
is the time
of maximum brightness, which is well constrained, as opposed to the longer-lasting
time of minimum or
mean light. To predict the time of maximum one adopts an ephemeris of the
form

\begin{equation}
\label{ephem}
C = E_0 + P_0 N,
\end{equation}

\noindent
where $E_0$ is an adopted origin or epoch, $P_0$ is the period at $E_0$ and $N$ is the
number of
cycles elapsed between $E_0$ and $C$. An initial estimate of the number of cycles
between the observed time of maximum $O$ and the reference $E_0$ is simply,

\begin{equation}
\label{cycles}
N = \left \lfloor \frac{(O - E_0)}{P_0} \right \rfloor,
\end{equation}

\noindent
where the incomplete brackets indicate the rounding down to the
nearest integer. However, we must note that if the time between $E_0$ and the observed
time of maximum
$O$
is much larger than the period, and the period change rate is large enough, the $O-C$
difference can exceed one or more cycles and then one must correct for these extra
cycles to obtain a correct $O-C$ diagram. This exercise may prove difficult 
if there are large gaps in the time-series but is rather straightforward otherwise, as
is the present case of M5.

A plot of the number of cycles $N$ vs $O-C$, usually referred to as an $O-C$ 
diagram, and its
appearance evidence a secular variation of the period (mostly parabolic but a higher
order is also possible) or the fact that the period $P_0$ used in the ephemerides  is
wrong, which produces a linear $O-C$ diagram. 
 
Let us assume the intitial model a cubic distribution of the $O-C$ residuals as
function of time, represented by the number of cycles $N$ elapsed relative to the
initial epoch $E_0$. The linear and quadratic cases are then particular solutions
of this more general representation:

\begin{equation}
\label{parab}
O-C = A_0 + A_1 N + A_2 N^2 + A_3 N^3,
\end{equation}

\noindent 
or,

\begin{equation}
\label{model}
O= (E_0 + A_0) + (P_0 + A_1) N + A_2 N^2 + A_3 N^3.
\end{equation}

From eq.\ref{model} it is clear that $E_0 + A_0$ is the corrected epoch for a given
fit and that $P_0 + A_1$ is the corrected period at that epoch.

Taking the derivative allows us to calculate the period at any given $N$,

\begin{equation}
\label{PdeE}
P(N)= \frac{dO}{dN} = (A_1 + P_0)  + 2 A_2 N + 3 A_3 N^2,
\end{equation}

\noindent
and,

\begin{equation}
\label{Pdot}
\beta \equiv \dot{P}=\frac{dP}{dt}= 2 A_2 \frac{dN}{dt} + 6 A_3 N \frac{dN}{dt}.
\end{equation}

Since the time elapsed to a given epoch is the period times the number of
cycles, $t=PN$, then $\beta$ at a given epoch $N$ and its secular variation
$\dot{\beta}$ can be written respectively as

\begin{equation}
\label{BETA_E}
\beta =\frac{2 A_2}{P} + \frac{6A_3 N}{P},
\end{equation}

\noindent
and

\begin{equation}
\label{BETAdot}
\dot{\beta}=\frac{6A_3}{P^2} -\frac{\beta}{P^2}(2A_2+6A_3N).
\end{equation}

Eq. \ref{BETAdot} represents the variation of period change rate at a given time $N$.
For $N=0$ and $P=P_0$ we get,

\begin{equation}
\label{BETA_0}
\beta=\beta_0=\frac{2A_2}{P_0}, 
\end{equation}

\noindent
and

\begin{equation}
\label{BETAdot0}
\dot{\beta_0}=\frac{6A_3}{P_0^2} -\frac{4A_2^2}{P_0^3}.
\end{equation}

If we set $A_2=A_3=0$ in eq. \ref{model} we would be dealing with the
linear case were the original estimate of the period $P_0$ should be corrected to
$P_0 + A_1$.

We have used the available data in the literature since 1889 to build 
up a collection of as many times of maximum light of as many RRLs as
possible. Some authors have already published times of maximum for some variables in
M5 (e.g. Coutts 1971, Kukarkin \& Kukarkina 1971), and in these cases 
we have adopted
their values. For other authors, we have used the published
photometric data to estimate the times of maximum.  When necessary we have
transformed Julian days into heliocentric Julian days. The complete collection of
the times of maximum for each star
is given in Table \ref{tab:maxima}, of which we show here just a small
portion since the
full table is only available in electronic form. To calculate the $O-C$
residuals as
described above, we adopted for each studied variable the ephemerides given by AF16 in
their table 3, and also listed in columns 3 and 4 of the present Table
\ref{variables}.

The resulting $O-C$ diagrams are shown in Fig. \ref{diagsOC} for every
variable with
observed times of maximum. For stars numbered above V100 we have not found
published photometry in the literature, thus the times of maximum light reported in
Table
\ref{tab:maxima} are those that we have calculated from the data available in AF16;
although this limitation inhibits any period change analysis at present, we
publish those times of maximum for the benefit of any possible future analysis.
Therefore, Table \ref{variables} and Fig. \ref{diagsOC} include only up to variable
V100.

\begin{figure*}
\includegraphics[width=17.0cm,height=20.cm]{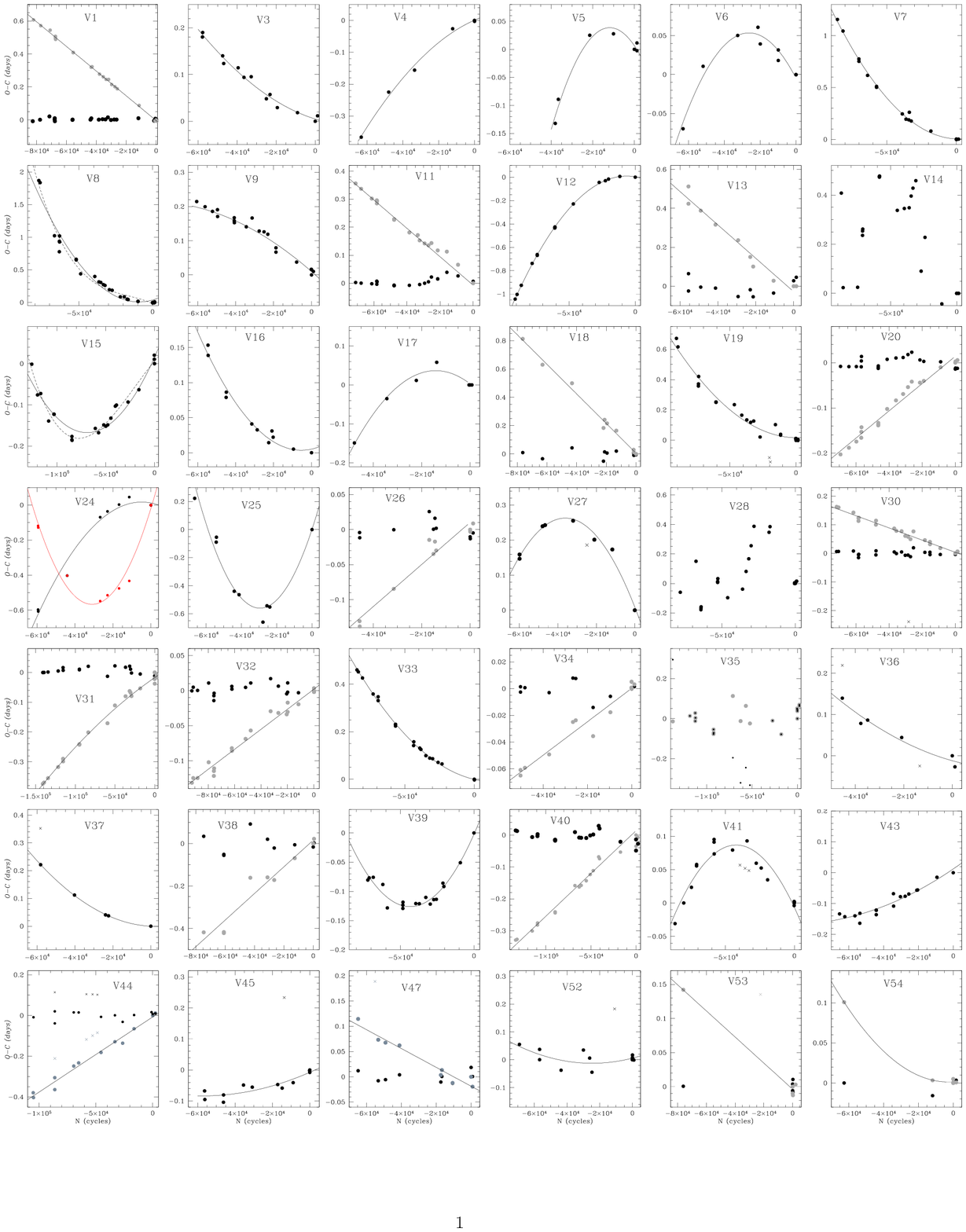}
\caption{$O-C$ diagrams. For stars with linear distributions we have plotted with gray
symbols the distribution using the ephemerides listed by AF16 (columns 3 and 4 in
Table \ref{variables}) and with black symbols the
distribution after the refined ephemerides (columns 5 and 6 in Table \ref{variables}).
Those $O-C$ residuals plotted with an X symbol were not considered in the fits.}
   \label{diagsOC}
\end{figure*}

\begin{figure*} 
\addtocounter{figure}{-1}
\includegraphics[width=17.5cm,height=17.5cm]{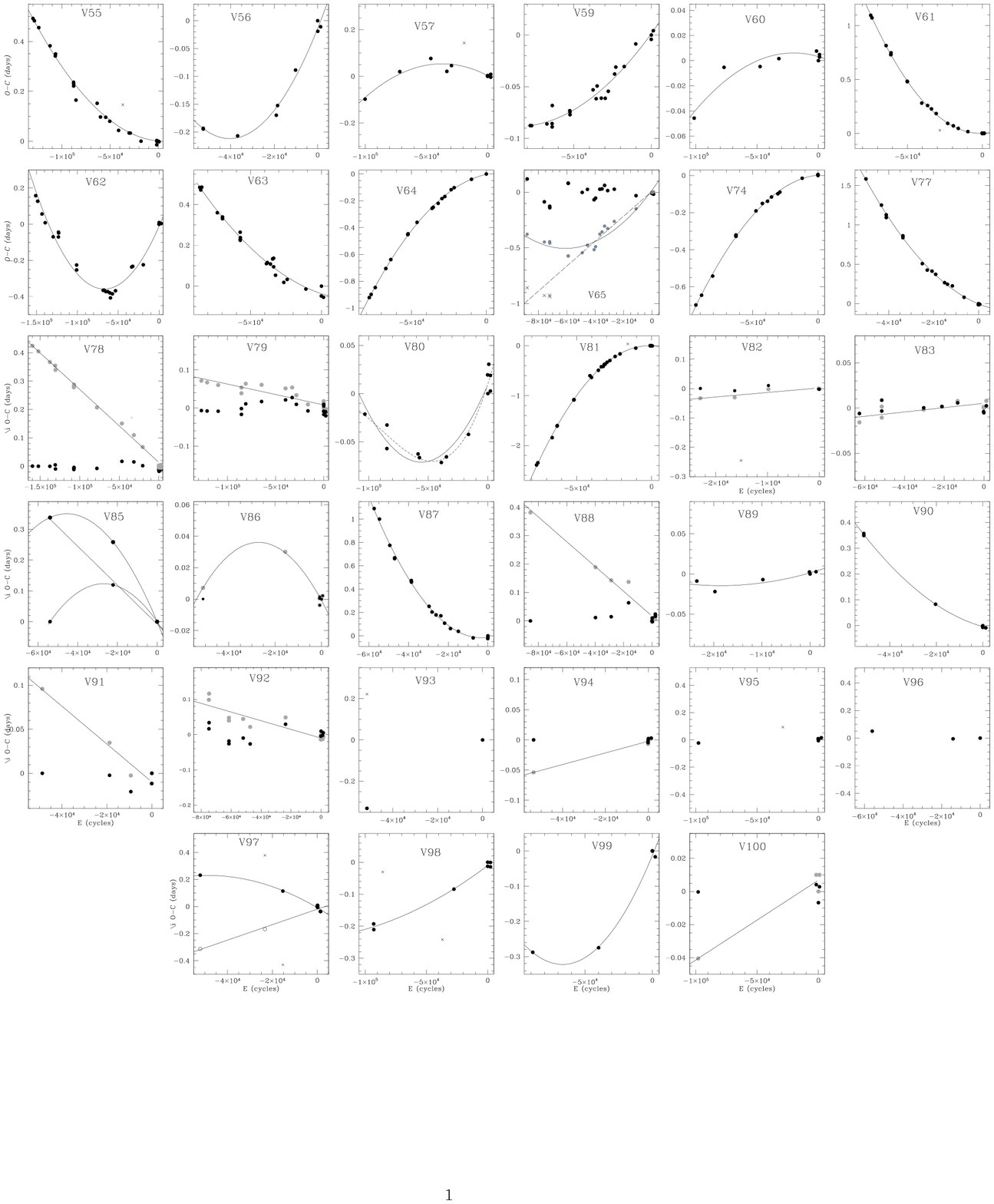}
\caption{Continued}
\end{figure*}

\subsection{The $(P,\beta)$ grid approach}
\label{PBeta}

When studying the period variations of variable stars it is common that the
available data do not span long enough or that the number of times of maximum light is
small, making it difficult to approach 
the problem via the classical $O-C$ residuals analysis.
Alternatively, we have recently used  recently the approach
introduced by Kains et al. (2015) to calculate $\beta$
for a sample of RRLs in NGC 4590 and also employed for NGC 6229 (Arellano
Ferro et al. 2015b); we briefly summarise this method here. The period
change can be represented as, 

\begin{equation}
 P(t)=P_0+\beta(t-E),
\end{equation}

\noindent
where $\beta$ is the period change rate, and $P_0$
is the period at the epoch $E$. The number of cycles at time $t$
elapsed since the epoch $E$ is:

\begin{equation}
\label{eq:NE}
N_E = \int_{E}^{t} \frac{dx}{P(x)} = \frac{1}{\beta} {\rm ln}[ 1 +
\frac{\beta}{P_0}(t-E)],
\end{equation}

\noindent
hence the phase at time $t$ can be calculated as

\begin{equation}
\label{eq:chanceperiod}
 \phi(t)=N_E-{\left \lfloor N_E \right \rfloor}.
\end{equation}

Then we construct a grid $(P,\beta)$ with values in steps of $10^{-6}d$ and 0.01
d Myr$^{-1}$ in $P$ and $\beta$ respectively, around a given starting period $P_0$, 
and examine the light curve dispersions for each pair. We choose as a solution the
one that produces the minimum dispersion of all the data sets phased simultaneously.
Since the penalty for increasing dispersion depends on the number of data, this method
works best for variable stars with many observations, as it can otherwise find many
local minima in the grid. We also note that when a solution with $\beta$=0 was among
the 10 light curves with the least dispersion, we adopted this solution, corresponding
to a simple period adjustment.

A comparison and a discussion of the grid results with the ones from the
$O-C$ method and those obtained by SZ11 are given in $\S$ \ref{sec:IND_STARS}.

\subsection{Phasing of the light curves with secular period change}
\label{phasing}

\begin{figure*}
\includegraphics[width=17.0cm,height=8.cm]{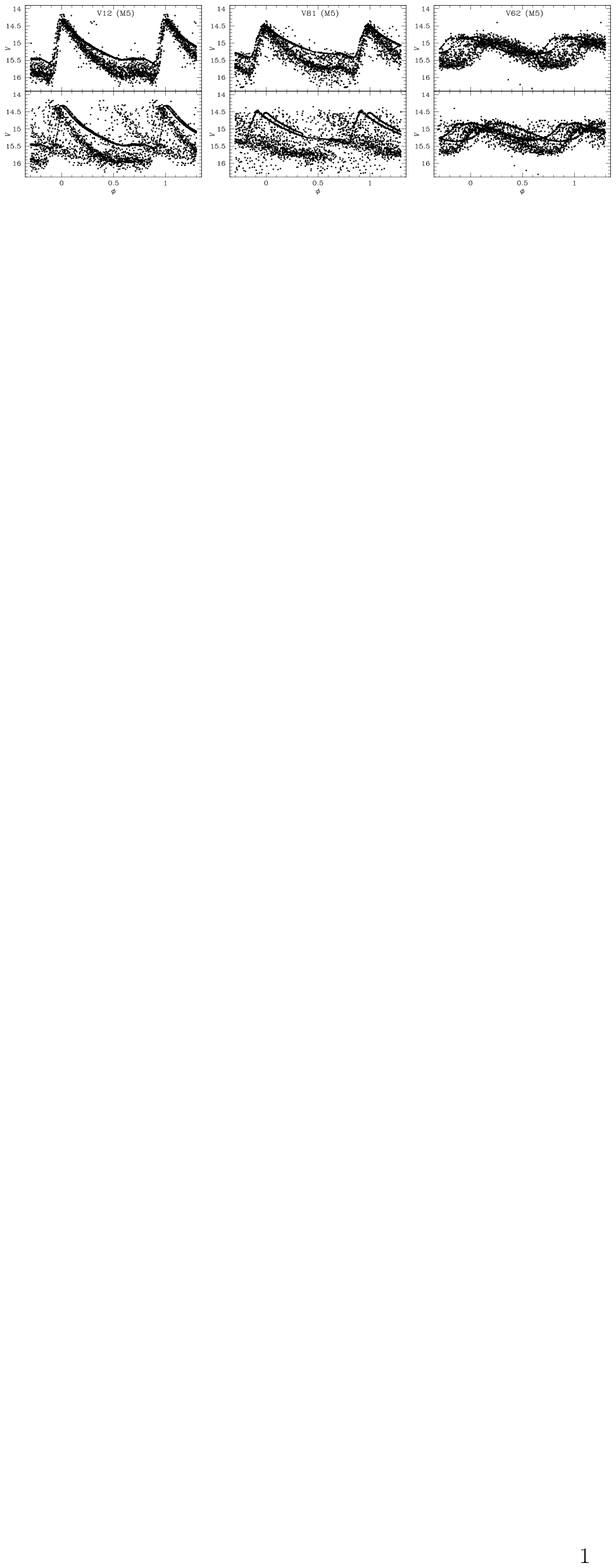}
\caption{Three selected examples of light curve phasing with ephemerides including the
secular period variation parameters described in $\S$ \ref{OmC} and given in Table
\ref{variables} (top panels). In the bottom panels the light curves were phased with a
constant period ephemerides, which makes evident the lack of proper phasing of the
several data sets involved (see $\S$ \ref{phasing} for a discussion).}
   \label{3examples}
\end{figure*}

Once a period change rate has been calculated, a natural test is to phase
the light curve given the new ephemeris. For the sake of brevity we do not plot all
the stars
in Table \ref{variables} but show three representative examples in Fig.
\ref{3examples}. In the lower panels we show the light curve of all available data
phases with a constant period ephemeris (columns 3 and 4 in Table \ref{variables}).
The upper panels show the light curve phase with the period change ephemerides
(columns 5, 6 and 7 in Table \ref{variables}) according to eqs.  \ref{model},
\ref{PdeE} and \ref{Pdot}. V12 is an RRab star for which the value of $\beta$ found by
the $O-C$ and the grid method agree and the new phasing is correct. For the RRab V81
the grid method failed in finding an optimal solution. The $O-C$ approach finds a
period change rate that phases well the available data. For the RRc
V62 the grid method did not converge and although the $\beta$ value from the $O-C$
approach improves the light curve phasing, it does not produce a perfectly phased
light curve for all the data sets. We believe this is a reflection of the scatter and
uncertainties inherent to the $O-C$ diagram. 
The grid method is very sensitive to the dispersion of the light curves involved
in the calculation and hence in some instances we could not derive a convincing
value of $\beta$. In Table \ref{variables} we mark with a colon those values of
$\beta$ that we consider to be particularly uncertain.

\subsection{Discussion on the period changes and individual cases}
\label{sec:IND_STARS}

From the $O-C$ diagrams in Fig.\ref{diagsOC} it is clear that the majority of the
stars display either a linear or a quadratic distribution. The linear cases imply a
constant period whose appropriate value is estimated by requiring that the slope of
the line be zero. The refined ephemerides are given in columns 5 and 6 of Table
\ref{variables} as $E_0+A_0$ and $P_0+A_1$ (see eq. \ref{model}). These linear
cases are all labelled $b$ in the last column of Table \ref{variables} and for
homogeneity we calculated their $\beta$ by adjusting a parabola to the $O-C$
distribution; as can be seen in all these cases $\beta$ is not significantly
different from zero, hence we conclude that their period has remained constant for
the last century. For a few stars whose apparent linear solution depends on a
very low number of points, we have adopted $\beta$=0, but have refrained from
performing any
further statistics (e.g. V53, V94, V95, V96 and V100). Similarly, some quadratic
solutions based on a low number of epochs are listed in Table \ref{variables} as a
reference but they are not highlighted since a solution can easily change by adopting
a slightly different number of cycles, see the explicit examples of V85 and V97 in
Fig. \ref{diagsOC}.
The boldface highlighted values of $\beta$ in Table \ref{variables} are our
recommended values as they are better estimated and produce a good
phasing of
the combined light curves. The  uncertainties given in parentheses after the
$(O-C)$ results were calculated from the
uncertainty of coefficient $A_2$ following eq. \ref{BETA_0} and neglecting the
period uncertainty. The resulting $P$ and $\beta$ from the 
grid approach ($\S$ \ref{PBeta}) are listed in column 8. The $(P,\beta)$ method
is a useful independent check on the values found with the $(O-C)$ method. However,
because it relies on the comparison of the string length of phased light curves,
rather than on a
standard goodness-of-fit statistic, it is difficult to obtain rigorous estimates of
the uncertainties associated with the values of $\beta$ yielded by the $(P,\beta)$
method. Such estimates are possible, but require detailed Monte Carlo simulations that
are beyond the scope of this study.

Except for a few cases with an unclear solution, mostly due to the scarcity of times
of maximum light or, in three cases (V14, V28 and V35) due to an irregular
distribution of the $O-C$
residuals, all the stars studied here show either a
smooth period change or the period appears to be constant over the last
century. 
We found three cases where the $O-C$ residuals are
best represented by a cubic equation, implying a secular variation of the period
change rate $\beta$ (V8, V15 and V80). We do not confirm the irregular period
variations
reported by SZ11 in at least one third of the stars studied by both teams.
We attribute this to the fact that in long time baseline data sets and for stars with
short periods, like RRLs, the $O-C$ residuals can easily exceed one or two
cycles and then the number of cycles might be miscounted. We have paid
special attention to the proper counting of cycles elapsed since the reference epoch
in the adopted ephemerides and we could always, or almost always,
make the
residuals to be consistent with a linear or a parabolic distribution, which in turn
are
easier to understand on evolutionary grounds than the stochastic oscillations in the
$O-C$ diagrams.

In what follows we compare the resulting values of $\beta$ obtained from the two 
approaches described in $\S$ \ref{OmC}
and $\S$ \ref{PBeta}, and subsequently with the results of SZ11. Finally, we will 
highlight and comment on some individual peculiar cases.

\subsubsection{$O-C$ vs $(P,\beta)$}
\label{OC-PB}

\begin{figure}
\includegraphics[width=7.cm,height=11.cm]{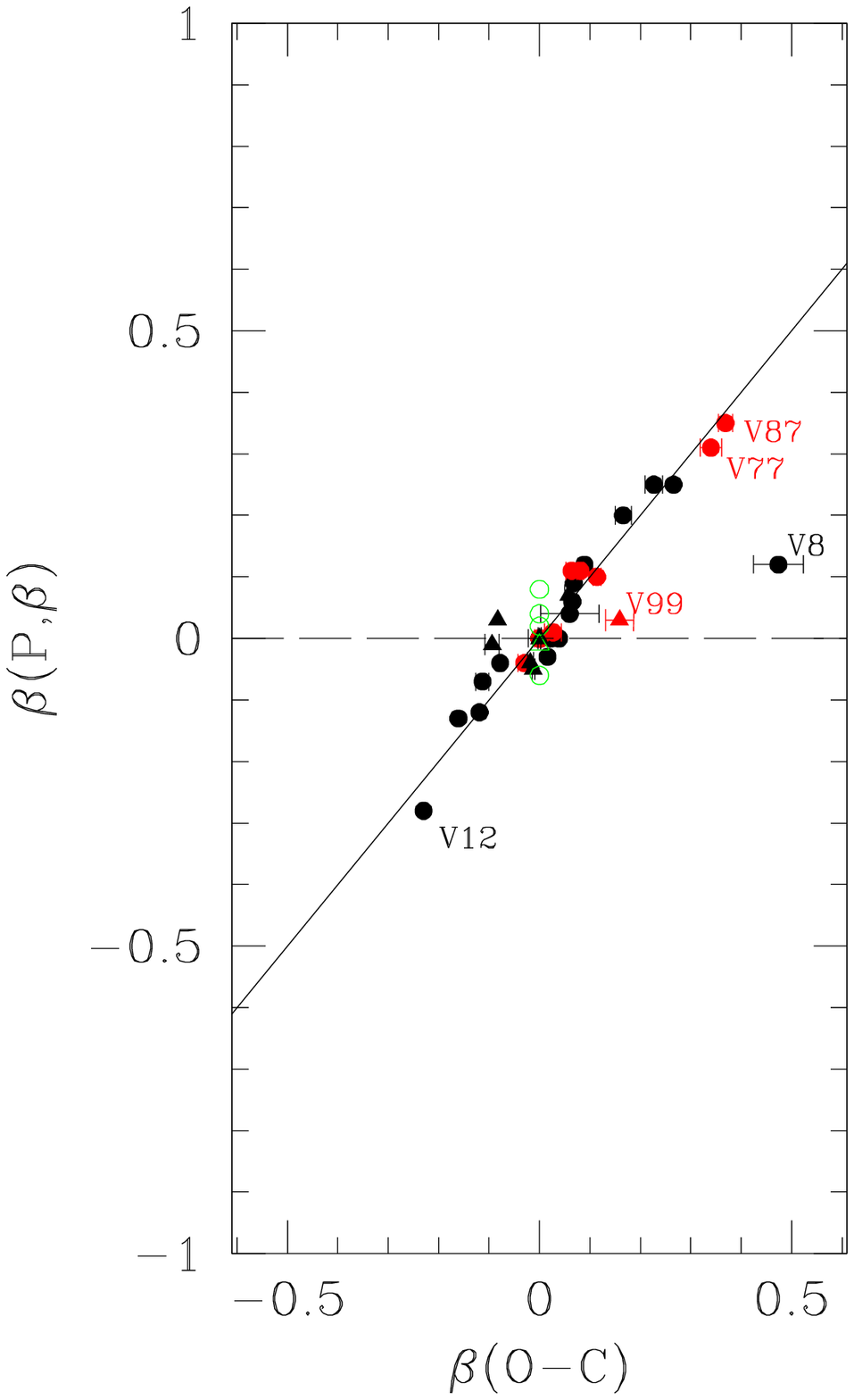}
\caption{Comparison of period change rates $\beta$ (in d~Myr$^{-1}$) from the 
$O-C$ approach described in $\S$ \ref{OmC} and the grid $(P,\beta)$ method of $\S$
\ref{PBeta}. Circles and triangles represent RRab and RRc stars respectively. Red
circles represent evolved stars identified by AF16 (see text for a discussion).
Error bars on $\beta (O-C)$ values are drawn only when they are noticeably bigger
than the symbol size.
Open green symbols are those stars with a constant period ($\beta = 0$) according to
the $O-C$ approach, but probably in most cases due to the presence of strong Blazhko
and/or scattered and scarce data, a non-zero value of $\beta$ was found by the grid
$(P,\beta)$
approach. See $\S$ \ref{sec:IND_STARS} for a discussion of individual stars.}
   \label{BocBpb}
\end{figure}

In Fig. \ref{BocBpb} the $\beta$ values from the $O-C$ and $(P,\beta)$ approaches
are
compared.
We were able to estimate $\beta$ by the two independent methods for 38 stars. The
agreement of the
two methods is generally very good, within 0.05 d~Myr$^{-1}$ for 32 stars, between
0.08 and 0.13 d~Myr$^{-1}$ for 3 stars, and there are 3 outliers; V8  in
the plot plus V4 and V25 not shown in the figure. For 26 stars the $O-C$
diagrams
suggest a constant period which could be refined by fitting a straight line and then
calculating $P_0 + A_1$ (see eq. \ref{PdeE}), these have a label $b$ in the notes
column of Table \ref{variables}. For some of these stars (V1, V13, V20, V40 and V96)
the
$(P,\beta)$ approach also suggested a small period change rate. These
stars are plotted with green symbols in Fig. \ref{BocBpb}. We call attention to
those stars with discrepant values of $\beta$ from the two approaches (V82, V85,
V88, V95, V97, V99). These stars either show prominent Blazhko amplitude modulations
(AF16), and/or have very few epochs (Fig. \ref{diagsOC}) or have a strong bump near
the maximum light (V95, V99); see their light curves in AF16); all these circumstances
act against a good determination of the period change rate. We have not plotted in
Fig. \ref{diagsOC} those stars with sparse data. Further comments on these stars are
found in $\S$ \ref{IndCases}.

\subsubsection{Comparison with {\rm SZ11}}
\label{OC-SZ11}

An overall comparison with the results of SZ11 and a discussion of some specific 
discrepant cases are of interest and they will be addressed below. The comparison is
graphically displayed in Fig. \ref{BvsB}.

There are 80 stars in common in Table \ref{variables} and SZ11. For 27 of them our
$\beta$
values agree within
0.03 d~Myr$^{-1}$ and for 7 the differences are between 0.05 and 0.18 d~Myr$^{-1}$. 
These cases are plotted with filled circles in Fig. \ref{BvsB}. We also note that
for 13 stars for which our $O-C$ approach suggested a clear period variation, SZ11 do
not report a value of $\beta$
generally because their $O-C$ diagrams show an irregular variation. We have found that
in the majority of these cases a proper counting of cycles leads to a homogeneous
parabolic $O-C$ distribution and hence to a clear value of $\beta$. To show the range
of $\beta$ implied by these stars, we have plotted them as open symbols in Fig.
\ref{BvsB}. Likewise for 7 stars we found a linear $O-C$ distribution instead of the
irregular one of SZ11.

For 10 stars, both SZ11 and us found no significant secular period variations, i.e.
$\beta \sim 0.$ These
are not represented in Fig. \ref{BvsB} and most of them are the ones with straight
$O-C$ residuals distribution in Fig. \ref{diagsOC}.

\subsubsection{Comments on individual peculiar cases}
\label{IndCases}

The following individual cases deserve specific comments:

\begin{figure}
\includegraphics[width=7.cm,height=11.cm]{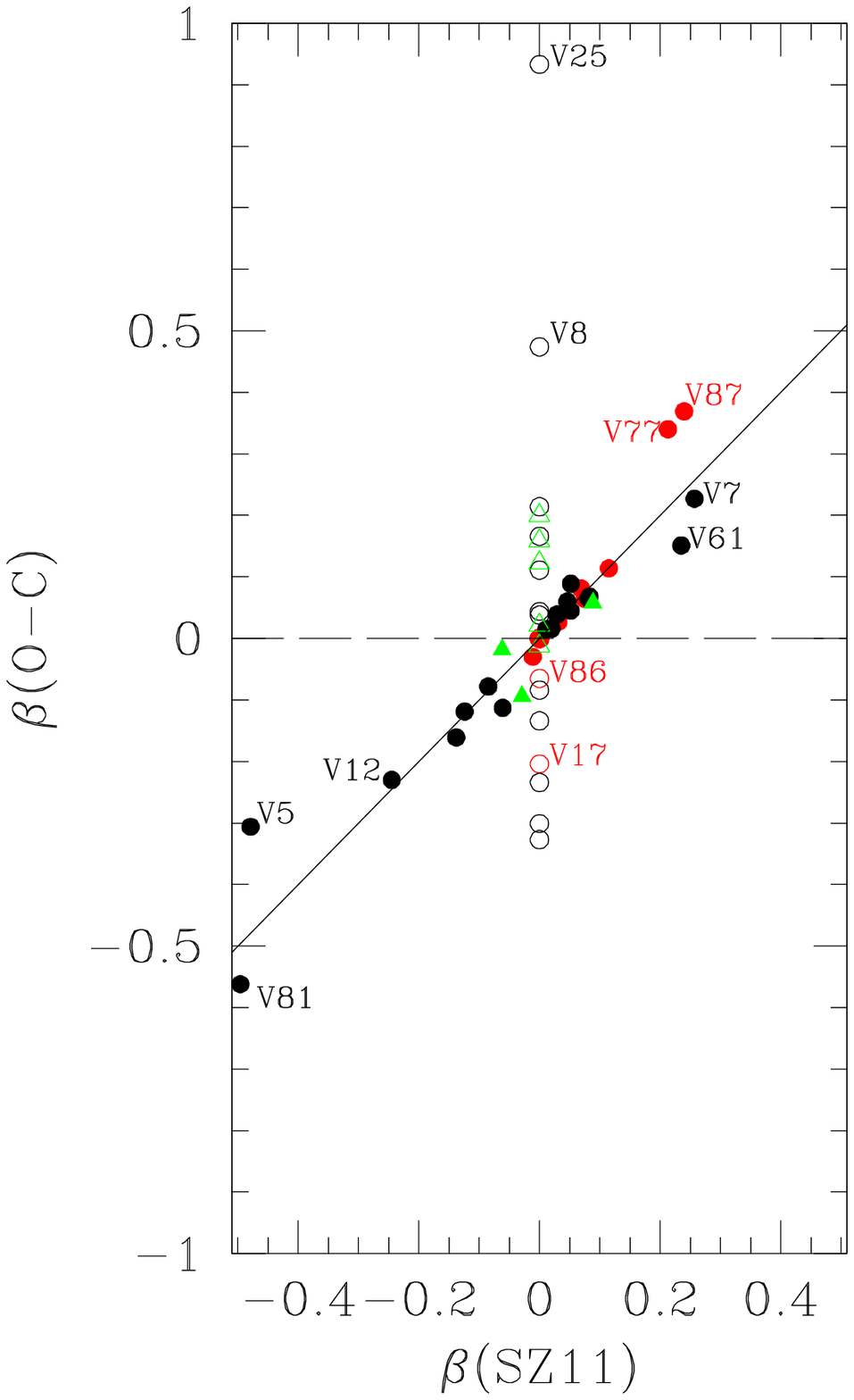}
\caption{Comparison of period change rates $\beta$ from SZ11 and the $O-C$ approach in
the present paper for stars in common. Circles and triangles represent RRab and RRc
stars respectively.
Red circles represent evolved stars according to the amplitude-period diagram of AF16 
(see $\S$ \ref{EVOL_HB} for a discussion).  Open symbols are used for star with a
value of $\beta$ in the present work but not calculated by SZ11, mostly because they
were found by SZ11 with a peculiar $O-C$ residuals distribution; they are plotted
at $\beta (\rm{SZ11})=0$.}
   \label{BvsB}
\end{figure}

\noindent
{\bf V4, V17, V19, V24, V25, V27, V52, V56, V62, V65, V80, V85, V98}. For these stars
SZ11 found a peculiar $O-C$ distribution.
Contrary to this we found a clear parabolic distribution indicating a smooth secular
period variation. We attribute this difference to the counting of cycles that, if
it is not properly made, may cause that the $O-C$ diagram can show peculiar
deformations. 

\noindent
{\bf V11, V18, V38, V40, V47, V88, V92}. These are the
stars with a linear distribution of the $O-C$ residuals in Fig. \ref{diagsOC} but
found with an irregular distribution by SZ11. 

\noindent
{\bf V1, V13, V20, V30, V32, V34, V78, V79, V83, V91}. For these stars both SZ11 and
this work agree that their period is not undergoing a secular variation.

\noindent
{\bf V8, V15, V80}. These stars show a period increase in both the present work and in
SZ11. We found that their $O-C$ diagrams
admit a parabola or a third degree polynomial, which implies a variable period change
rate. A similar result is found by SZ11 for V8 but no value of $\beta$ is
reported. We were not able to find a good phasing for V8 with the $(P,\beta)$ approach
thus its value of $\beta$ in Table \ref{variables} is marked with a colon.

\noindent
{\bf V4, V25}. The very large values of $\beta$, $-2.31$d~Myr$^{-1}$ for V4 and
$+4.01$d~Myr$^{-1}$
for V25, determined by the $(P,\beta)$ method (not shown in
Fig. \ref{BocBpb}) are at odds with those obtained by the $O-C$ method. These
two stars show very clear and prominent Blazhko amplitude and phase modulations
(see the light curves in AF16) which make the determination of the period change, by
both the $(O-C)$ and $(P,\beta)$ methods very uncertain.

\noindent
{\bf V14, V35}. These are examples of irregular $O-C$ diagrams according to SZ11. Our
$O-C$ distribution confirms this. We found no way to reconcile the residuals with a
homogeneous variation. The RRab V14 is also a Blazhko variable and the RRc V35
displays a prominent
bump near maximum light which has probably influenced the historical determinations
of the times of
maximum light. Some evidence of Blazhko modulations have also been found in V35 (see
the light curves in AF16). 
Depending on the cycle counting, especially in data from around the 1960's, one could
interpret
the $O-C$ diagram as having a constant period, in agreement with the result of
SZ11, or as having an abrupt period change at that time (gray and black dots in Fig.
\ref{variables} respectively). It is interesting to note that the
$(P,\beta)$ approach suggests $\beta$=+1.750, a very large value which may be a
spurious result driven by the presence of the bump and the Blazhko modulations. Other
regions of the P-$\beta$ plane were explored in search of other possible solutions but
no convincing phasing was found.

\noindent
{\bf V24}. We found a $O-C$ distribution similar to the one of SZ11.  It can be
approximated by
an upward parabola (red in Fig. \ref{BvsB}), 
in this case the resulting value of $\beta$ (+0.868 d~Myr$^{-1}$) would be the largest
in
M5, similar to the case of V25. We note
however that through an appropriate count of cycles one could accomodate a downward
parabola (black in Fig. \ref{BvsB}) with a negative value of $\beta$ ($-0.327$
d~Myr$^{-1}$). The $(P,\beta)$ method failed to find a convergent solution for this
star.
Thus, we are not conclusive about the period variations of this star.

\noindent
{\bf V28}. As for V14 and V35 we could not reconcile our $O-C$ diagram with a
homogeneus period change but found an irregular
distribution. SZ11 found a neat $O-C$ diagram
that would be consistent with a period decrease, they reported $\beta$= -0.236
d~Myr$^{-1}$.
The $(P,\beta)$ approach leads to $\beta$= -0.16 d~Myr$^{-1}$ in reasonable agreement
with SZ11. V28 has a very strong amplitude Blazhko modulation (AF16) which apparently
has altered the times of maximum sufficiently to produce the irregular $O-C$ diagram. 

\noindent
{\bf V31}. Although this star was found by SZ11 to have a constant period, we found
that a very gentle downward parabola can be admitted, as opposed to a tilted
straight line, which corresponds to a mild period decrease with $\beta =-0.012$
d~Myr$^{-1}$.

\noindent
{\bf V36}. No period variation was found by SZ11. Our $O-C$ diagram might also admit
a small period increase with $\beta =+0.03$ d~Myr$^{-1}$ represented by a soft upward
parabola.

\noindent
{\bf V38}. This apparently constant period star has a scattered $O-C$ diagram probably
due to its Blazhko nature detected by Jurcsik et al. (2011).

\noindent
{\bf V41}. We, like SZ11, find a downward parabola and a very similar value of
$\beta$. However we note that we had to ignore three maxima from Kukarkin \& Kukarkina
(1971) that seemed discordant.

\noindent
{\bf V62, V65}. Both $O-C$ diagrams of SZ11 and ours in Fig \ref{BvsB} show small
discrepancies about the parabola of these otherwise period-increasing stars. SZ11 do
not report a value of $\beta$. In fact V65 could also be interpreted as a constant
period variable if the four oldest data points were shifted one cycle downwards
(segmented line in Fig. \ref{diagsOC}).

\noindent
{\bf V53, V54, V85, V86, V92-V100.} These stars have very few observed times of
maximum
light, and consequently the $O-C$ solutions are not always clear and often depend on
one or two points, e.g. see the cases of V85 and V86 where the removal of one datum
can change
the
solution from quadratic to linear. For V92 and V97 we were unable to find a proper
phasing of the data by the $(P,\beta)$ approach and the reported $\beta$ are marked
with a colon. In Table
\ref{variables} and Fig. \ref{diagsOC} we have suggested likely solutions for these
stars, however
inconsistencies with SZ11 should not be either surprising or significant. These
stars
will not be included in the overall period change statistics of the cluster. V53 and
V86 were not included by SZ11.

\noindent
We note that despite these
adverse circumstances one of the competing solutions produce a reasonable
phasing of the light curve, as can be seen in Fig.

\begin{figure*}
\includegraphics[width=17.0cm,height=10.cm]{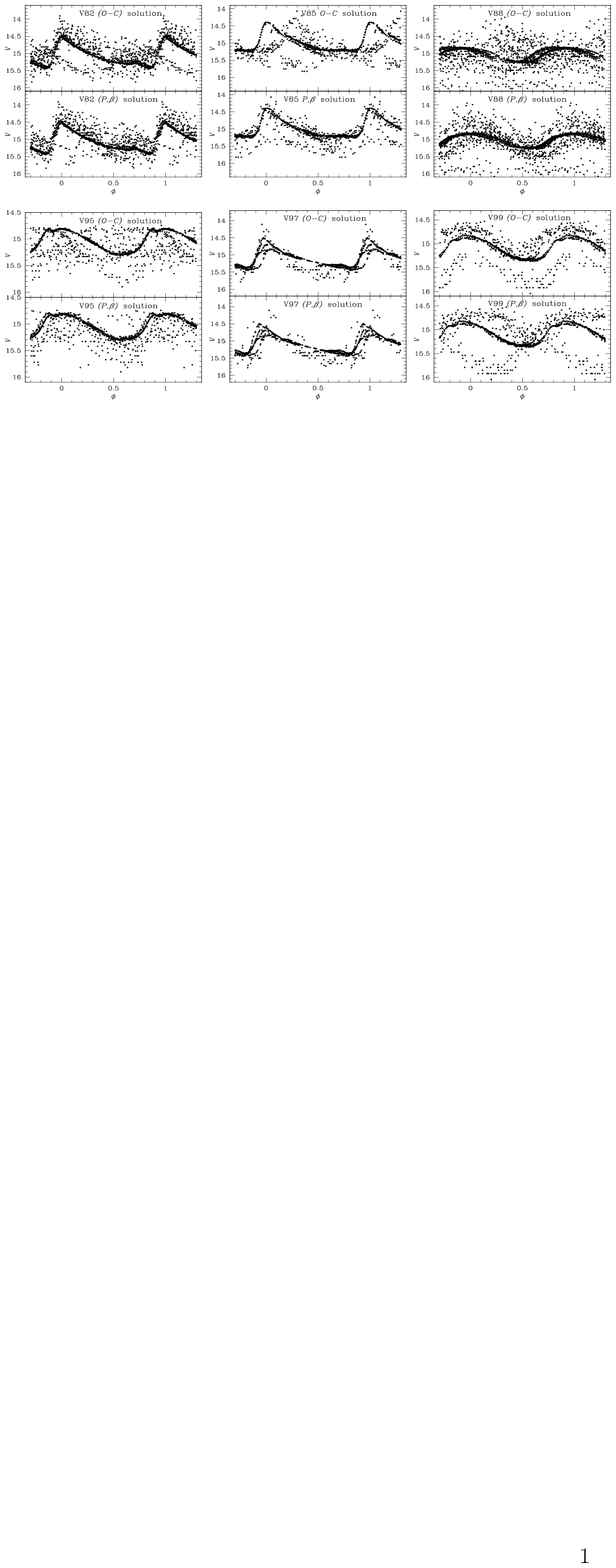}
\caption{Phasing of the light curves of stars with discrepant values of $\beta$
values from the $(O-C)$ (top panels) and the $(P,\beta)$ (bottom panels) approaches.
One of the competing solutions seems to produce a proper phasing of the light curve.
 Refer to $\S$ \ref{IndCases} for a discussion.}
   \label{examples}
\end{figure*}

Further comments on V82, V85, V88, V95, V97, V99. It has been pointed out in 
$\S$ \ref{OC-PB} that data on these stars are sparse, and that they show either
prominent Blazhko modulations and/or a bump near the maximum light (AF16), hence it is
difficult to be conclusive on their secular period behaviour. In Fig.
\ref{examples} the light curves phased with the two competing results are displayed.
In spite that one of the solutions produces a reasonable phasing, we warn that these
cases
should be considered as marginal until more data become available.

\section{Period changes and evolution in the HB}
\label{EVOL_HB}

It is clear that as stars evolve across the instability strip, their pulsation
period should increase if evolution is to the red and decrease if evolution is to
the blue.
However, it is also understood now that evolution is most likely not the only 
reason for period variations and stochastic variations have also been proposed
(Balazs-Detre \& Detre 1965). In fact several authors have found irregular and
complicated $O-C$ diagrams (e.g. SZ11, Jurcsik et al. 2001), which indeed would be
difficult to reconcile
with evolution exclusively. Sweigart \& Renzini (1979) suggested that mixing 
events in the core of a star at the HB
may alter the hydrostatic structure and pulsation period and
that a small transfer of helium inwards at the semiconvective zone, near the
convective core boundary, can be responsible for period changes, either positive,
negative, or irregular. However, we note that at least in the case of M5,
there is no need to claim for irregular period variations since an
improper counting of cycles,
particularly in long time baseline sets of times of maximum light, may be responsible
for
those apparent irregularities, with the probable exception of V14 and V35. 

It has been in the interest of several authors in the past to infer evolutionary
properties accross the instability strip by searching for preferential directions
of evolution as indicated by the secular period changes in RRLs. A
preferential period variation through the IS, in either direction, seems
however to be improbable. There are several reasons for this. Growing
evidence of several stellar populations coexisting in a given globular
cluster have been put forward in recent years, both theoretically and observationally
as well as on spectroscopic (Carretta et al. 2010; Gratton et al.
2012) and photometric (e.g. Piotto et al. 2007,  Sbordone et al. 2011, Alonso-Garc\'ia
et al. 2015) grounds. The presence of multiple populations, and their role
in the structure of the IS at the horizontal branch, have recently been
discussed by Jang et al. (2014) and Jang \& Lee (2015) in connection with
the origin of the Oosterhoff types dichotomy. They conclude that RRab-type stars in
OoII clusters belong to a second generation with the helium abundance $Y$ slightly
increased and carbon, nitrogen, and oxigen (CNO)
noticeably enhanced which produce more luminous and redder stars, therefore with
longer periods; in OoI clusters the RRab stars are members of the less luminous
first generation. However, this scenario, that aims to explain the Oosterhoff
dichotomy,
does not account for the non-preferential distribution of increasing and decreasing
periods in a
given globular cluster. 
Thorough and extensive investigations on the secular period change rates, 
$\beta$, in a 
few clusters have been carried out and the averages are not significantly different
from zero;
for M3 Corwin \& Carney (2001) found an overall average of +0.004$\pm$0.335 for 37
RRLs and also Jurcsik et al. (2012) found a small positive value of about
+0.01 for some 54 RRLs. For M15
Silbermann \& Smith (1995) found an average of +0.06$\pm$0.24 for 27 RRLs and
excluding the double mode or RRd pulsators also Smith \& Wesselink (1977) found an
average of +0.11$\pm$0.36; for NGC 4590 Kains et al. (2015) found an average of 
+0.02$\pm$0.57 for 14 stars;
NGC 7006 Wehlau, Slawson \& Nemec
(1999) found +0.03$\pm$0.14 for stars; for M28 Wehlau et al. (1986) found
-0.06$\pm$0.29 for 11 stars; for M14 Wehlau \& Froelich (1994) found +0.04$\pm$0.43
for 35 stars. The case with the largest positive average $\beta$ is $\omega$ Cen
(Jurcsik et al. 2001 their table 6) for which we calculate +0.170$\pm$0.561. All
the above cited uncertainties correspond to the standard deviation of the mean.
The case of IC4499 (Kunder et al. 2011) comes a bit as a surprise since the
average +0.29$\pm$0.60 from 39 stars (their Table 1 without three extreme cases) is
found despite of not having a very blue HB ($\cal L$=+0.11).
Also, in all the above studies no significant differences were found in the average
values of $\beta$ for the population of RRab and RRc stars.

On theoretical grounds, Lee (1991) demonstrated that in globular clusters with
blue HB morphologies, the average rate of period change $\beta$ would become
increasingly large but it will be small on average for the RRLs in 
globular clusters with reddest HB's (see his figure 4). An updated version of this
correlation,
provided by Catelan \& Smith (2015) (their figure 6.16), supplemented with new
theoretical calculations and more empirical results confirms the trend. 
However, on empirical grounds the trend of increasing $\beta$ with very blue HB types
is driven primarily by $\omega$ Cen [a
peculiar cluster
with a large star-to-star spread of heavy element abundances as pointed out by Catelan
\& Smith (2015) and Valcarce \& Catelan (2011)], but also by M3 and M28, and (to some
extent) M22. Further studies of secular period variations in globular cluters with
(very) blue HB types are strongly encouraged, as this would enable a better
investigation of the possible connection between period change rates and evolution
along the HB.

For M5 several authors have concluded that the averages of period changes do not
show significantly preferential increasing
or decreasing periods (e.g. Reid 1996, Coutts
\& Sawyer-Hogg 1969, SZ11). SZ11 found an overall average of
-0.006$\pm$0.162 for 55 stars. 
From our Table \ref{variables} we can see that out of the 76 stars studied, constant
periods are evident in 18 RRab and 8
RRc while parabolic or cubic secular variations are observed in 37 RRab and 9 RRc
stars. Of the remaining four stars, in one (V93) the solution
depends on one point, and the other three are irregular (V14, V28 and V35). Thus, 66\%
of the
studied population has secular period variations, and 
34\% shows periods that seem to have been stable over the last century. According
to the $O-C$
results in Table \ref{variables}, we see that among the RRab stars 23 have
$\beta
> 0$, 15 have $\beta < 0$ and 19 have $\beta = 0$ for a total average
+0.026$\pm$0.210. 
For the RRc stars 6 have $\beta > 0$, 4 $\beta < 0$ and 8 $\beta = 0$ for an average
of +0.023$\pm$0.073. Thus, we find small positive average values for $\beta$ in
the RRLs of M5 and there seems to be a small surplus of RRab stars with increasing
periods.

Furthermore, while it is true that other sources of period change noise, such as
stochastic processes, and/or with mixing events in the stellar core, are superposed
on evolutionary period variations, it is of fundamental interest to isolate stars with
evidence that suggests they are in an advanced stage of evolution, and to determine
whether they
are subject of systematic larger positive period change rates.
Based on their distribution on the log $P - A_V$ plane, AF16 identified a dozen
RRab stars likely to be in an advanced evolutionary stage and found them consistently
more luminous than the rest of the RRLs. These are the stars V9, V11, V16, V17,
V20, V26, V39, V45, V77, V86, V87 and V90 and are ploted in Figs. \ref{BocBpb} and 
\ref{BvsB} with red colour. 
We can see that the agreement between the values of $\beta$ found in this work and
those of SZ11 is very good, except for
V17 and
V86 that were not included by SZ11. Also we note from Figs. \ref{BocBpb} and
\ref{BvsB}, that there is a clear trend for positive values of $\beta$ in these
stars, as expected if they are indeed advanced in their evolution towards the
AGB.
Among these evolved stars, three stand out for their large positive $\beta$, V77
(+0.340
d~Myr$^{-1}$), V87 (+0.369 d~Myr$^{-1}$) and V90 (+0.114 d~Myr$^{-1}$) (confirmed
by the grid results), and are
the clearest cases of evolved stars displaying large increasing period rates. We
conclude that in this group of RRab stars their period increase
rate and their advanced evolutionary stage are connected.

On the other hand, the stars with most extreme positive values of $\beta$, V8 (+0.474
d~Myr$^{-1}$) and V25 (+0.933 d~Myr$^{-1}$)
were not noticed by AF16 as probable evolved stars, and although V25 has been
identified as being a blended star (Arellano Ferro et al. 2015a),
a circumstance that should not affect the $O-C$ diagram, the large period increase
rates are well established.
Other stars with large positive values of $\beta$ and without other evidences of
advanced evolution are V7 (+0.474 d~Myr$^{-1}$) and V61 (+0.266 d~Myr$^{-1}$). The
origin
of the period change in these stars may be more connected with stochastic
processes and/or with mixing events in the stellar core as described above.

The existence of large negative values of $\beta$ should not be surprising.
Depending on the mass and chemical
composition on the ZAHB, the evolutionary tracks can have blue loops into the
IS (e.g. Caputo et al. 1978; Jang \& Lee 2015) and enhancement of $Y$ and CNO also
causes
the evolutionary tracks to appear redder and more luminous (Jang \& Lee 2015) making
stars
with negative $\beta$ to appear as more evolved. We remind that, as 
discussed by Silva Aguirre et al. (2008), approximately 22\% of the pre-ZAHB stars
may fall in the IS and present RR Lyrae-like pulsations and about 76\% of them are
predicted to have negative $\beta$ values of the order of $-0.3$ d~Myr$^{-1}$ although
more
extreme values, say inferior  than $-0.8$ d~Myr$^{-1}$, are possible.

Employing the grid of HB evolutionary tracks 
of Dorman (1992) for [Fe/H]=$-1.48$ and masses of 0.60 to 0.66 $M_{\odot}$, Jurcsik et
al. (2001) calculated period change rates accross the IS to be between -0.026 and
+0.745 d~Myr$^{-1}$. These values are in agreement with what we calculated for the
majority
of the RRLs stars in M5.
However, in cases with extreme values of positive or negative $\beta$, 
evolution may play only a partial role in the period secular variations, except
probably in stars with independent evidence that they are more advanced on their way
to the AGB.

There are examples in the literature of RRLs with very large period change rates
which would be difficult to explain exclusively by evolutionary arguments; the RRab
V104 (-19.796 d~Myr$^{-1}$), the RRc stars V160 (-10.18
d~Myr$^{-1}$) and 
V123 (+5.474 d~Myr$^{-1}$) and the RR Lyrae-like stars V48 (+15.428 d~Myr$^{-1}$) and
V92 (+13.941 d
/Myr)  in $\omega$ Cen (Jurcsik et al. 2001); the RRab star V29 (-17.8
d~Myr$^{-1}$)
in NGC 6981 (Bramich et al. 2016); the RRc stars V14 (+8.26 d~Myr$^{-1}$) and V18
(+4.0
d~Myr$^{-1}$) in NGC 6333 (Arellano Ferro et al. 2016b). In M5 we find V4
(–2.31 d~Myr$^{-1}$) and V25 (–0.93 d~Myr$^{-1}$). Other processes must be
at play in these cases.

It is thus clear that the RR Lyrae instability strip can be populated by
stars
of different ages, chemical compositions and evolving both to the blue or to the
red. This assertion, combined with the fact that stochastic processes related to the
semiconvention zone can also produce variations in the pulsation period,
indicate that no systematics trends in the period changes in a given cluster are to be
found.
Exceptions may be those globular clusters with very blue HB structures, according
to
theoretical predictions (Lee 1991, Catelan \& Smith 2015), and efforts towards detail
period change analysis in these clusters should be encouraged.
After these
considerations it is striking to note, however, that there are stars that manage to
mantain a stable period for over a hundred years: in M5 we identified 26 of them in
the sample of 76 stars studied, i.e. 34\%, and that a few stars in this
cluster with evidence of advanced evolution indeed show large positive period
change rates.

\section{Summary}
\label{summary}

Pulsation period changes have been analysed via the times of maximum light for 76 RR
Lyrae stars in M5. Archival data were collected from the literature that span up to 
118 years for many of the sample stars. No
signs of irregular or stochastic variations were found in the large majority of the
stars but
instead they have either a remarkably stable period or a secular
period change that can be
represented by a parabolic $O-C$ diagram. The only exceptions are V14, V28 and V35
for
which
their $O-C$ diagrams are irregular and we were unable to find a more simple solution.

The collection of times of maximum assembled in this work is published in electronic 
format for the sake of possible future analyses.

We have also used an alternative method first introduced by Kains et al. (2015) in
which a grid $(P,\beta)$  is built to allow the selection of the solution that
minimizes the dispersion
in the phased light curve. The values of $\beta$ found by the grid method
and those from the $O-C$ approach give very consistent results, generally within
$\pm$0.05 d~Myr$^{-1}$. Also the comparison with the
values of $\beta$ of SZ11 for the stars in common is very good.

Other than stellar evolution, the existence of stochastic effects influencing the
structure of the star in the IS may produce variations of the pulsational period.
This fact,
and the existence of multiple stellar generations with a range
in chemical compositions in the cluster, make that evolution accross the IS occurs
both to the red
and to the blue and that neither increasing nor decreasing periods are
significantly 
more frequent. This scenario is consistent with theoretical expectations for
globular clusters without blue HB morphologies. Most likely the observed period
variations are not the result of a single cause. Nevertheless, in M5 we have been able
to isolate a group of likely evolved  stars that show systematically positive, and in
some cases large, rates of period change.

\section*{Acknowledgments}
It is pleasure to thank Dan Bramich for his valuable input and numerous suggestions
and to Juan Carlos L\'opez B. for his help digitizing the bulky archival data
available since the late XIX century. Comments and suggestions from an anonymous
referee contributed to the improvement of certain passages and they are sincerely
appreciated.
JAA wishes to thank the Instituto de Astronom\'{\i}a
of the Universidad Nacional Aut\'{o}noma de M\'{e}xico
for warm hospitality during the preparation of
this work.
Finantial support from  DGAPA-UNAM, Mexico via grant
IN106615-17 and from CONACyT (Mexico) and Mincyt (Argentina) via the collaborative
program \# 188769 (MX/12/09) is fully acknowledged.
We have made an extensive use of the SIMBAD and ADS services, for which we
are thankful.

\end{document}